\documentclass[twocolumn,english,superscriptaddress]{revtex4-1}
\usepackage[T1]{fontenc}
\usepackage[latin9]{inputenc}
\usepackage{graphicx}
\usepackage{babel}
\usepackage{amsmath}
\usepackage[dvipsnames]{xcolor}

\begin{document}

\title{Periodic training of creeping solids}
\author{Daniel Hexner}
\affiliation{Department of Physics and The James Franck and Enrico Fermi
Institutes, University of Chicago, Chicago IL, 60637.}
\affiliation{Department of Physics and Astronomy,University of Pennsylvania, Philadelphia PA, 19104.}
\author{Andrea J. Liu}\affiliation{Department of Physics and Astronomy,University of Pennsylvania, Philadelphia PA, 19104.}
\author{Sidney R. Nagel}
\affiliation{Department of Physics and The James Franck and Enrico Fermi
Institutes, University of Chicago, Chicago IL, 60637.}

\begin{abstract}
We consider disordered solids in which the microscopic elements can deform plastically in response to stresses on them. We show that by driving the system periodically, this plasticity can be exploited to train in desired elastic properties, both in the global moduli and in local ``allosteric'' interactions. %where the strain at a local input is transmitted to a distant target site.  
Periodic driving can couple an applied "source" strain to a target strain over a path in the energy landscape. This coupling allows control of the system's response even at large
strains well into the nonlinear regime, where it can be difficult to achieve control simply by design. 
\end{abstract}
\maketitle

\section*{Introduction}
Metamaterials offer the possibility of creating a broad array of behaviors not found in ordinary, non-architected materials.  By manipulating the connectivity and strength of the structural units, rather than the composition or structure of the native material itself, metamaterials with unusual elastic response--mechanical metamaterials--can be designed.  These include systems that display phononic band gaps~\cite{kushwaha1993}, unusual structural deformations leading to negative Poisson's ratios~\cite{mullin2007pattern,bertoldi2017flexible}, negative-compressibility transitions~\cite{nicolaou2012mechanical}, topologically-protected modes~\cite{kane2014topological,chen2014nonlinear}, negative swelling~\cite{liu2016harnessing}, complex pattern formation~\cite{coulais2016combinatorial} and allostery-inspired responses where the imposed strain on a given site is propagated to a distant target~\cite{rocks2017designing,yan2017architecture,mitchell2016strain}. While such metamaterials can be designed and built on a relatively small scale, it is not always clear how to scale up the number of components or to control the microstructure at the microscopic level in order to achieve the desired behavior --  especially when the applied deformations are well outside the linear-response regime.

One successful approach has been to design the materials based on a repeating unit cell.  Such an approach requires precise control of each degree of freedom during fabrication of the unit cell, but has the advantage that each unit cell is the same.  
However, this strategy is unable to create a material with a heterogeneous or localized response, which by definition cannot be captured by identical responses in each unit cell.  A mechanical metamaterial with inhomogeneous response is difficult to design, since it requires detailed knowledge of structure and mechanics at the constituent scale and computer resources that grow with system size. It is also challenging to fabricate, since it requires control and manipulation at the constituent scale. 

The idea of ``directed aging''~\cite{pashine2019directed} circumvents these obstacles by starting with a disordered solid and training it while it ages by applying appropriate stresses in such a way that it ultimately evolves to have the desired functionality. Directed aging takes advantage of the natural tendency of a material to minimize its energy under stress by deforming plastically.
%Plastic deformation encodes memory stored in the material of the applied strains, as has also been proposed for selecting folding pathways in origami~\cite{stern2019learned}. 
A demonstrated process that can be understood in terms of directed aging is the heating of solid foams under pressure to create auxetic (\textit{i.e.}, negative Poisson's ratio) foams~\cite{lakes_r_Science_1987}. The concept of directed aging allows this process to be generalized to create materials with a variety of responses determined by the stresses applied~\cite{pashine2019directed}. Aging a system under a fixed shear stress, for example, leads to systems with high Poisson's ratios. The challenge for this general approach is to find appropriate flexible protocols for the training that will produce a broad class of response. 

In this paper we introduce a new strategy for directed aging. We show that systems that deform plastically and irreversibly via creep can be trained to develop complex responses by applying a \textit{periodic} training strain instead of a fixed training strain. To demonstrate the feasibility of this approach we start with the same model systems and train them to exhibit three different types of responses: negative Poisson's ratio, bi-stability, and mechanical allosteric response ~\cite{rocks2017designing,yan2017architecture}. We are able to control the responses far into the non-linear regime.

\section*{Model }

We start by approximating a disordered solid as a random spring network, where each
spring is defined by its spring constant, $k_{i}$, and its rest length
$\ell_{i,o}$. The elastic energy is quadratic in the deformation:

\[
U=\frac{1}{2}\sum_{i}k_{i}\left(\ell_{i}-\ell_{i,0}\right)^{2}.
\]
Our ensemble of spring networks in $d$ spatial dimensions is derived from packings of soft spheres
at force balance under an external pressure~\cite{Durian1995,Ohern,liu2010jamming}.
The centers of the spheres define the locations of the nodes, and
overlapping spheres are connected with springs. The rest length is
chosen to be the distance between nodes, so that in the absence of
any imposed deformation the system is unstressed and at zero energy. To eliminate surface effects we consider periodic boundary condition.
Using packings as a starting point for our metamaterials ensures
that systems are always rigid and allows the connectivity
of the network to be tuned by varying the external pressure on the original packings. 

We characterize the connectivity of the network by the coordination number $Z=\frac{2N_{b}}{N}$,
where $N_{b}$ is the number of bonds and $N$ the number of nodes.
At the jamming transition, where particles just touch, the coordination
number is the smallest possible needed to maintain rigidity, $Z_{c}$. In the large-system limit, $Z_{c}=2d$. Increasing
the pressure, $p$, increases $\Delta Z \equiv Z-Z_{c}$. %Whenever possible we work far from the jamming transition which is a more generic representation of a material. Nonetheless, in some cases we also exploit the anomalous elasticity near the jamming transition~\cite{ellenbroek2006critical}.

We include plasticity via the \textit{$\ell$-model} introduced in ~\cite{hexner2019nonlinear}. This model accounts for plasticity
via a change in the \textit{rest length} , $\ell_{i,0}$, of each bond $i$: each bond
changes its rest length to reduce its internal stress.
The rate of change of the length depends on the stress in the bond,
so that a bond elongates if it is under tension and shortens under
compression: 

\begin{equation}
\partial_{t}\ell_{i,0}=\gamma k_{i}\left(\ell_{i}-\ell_{i,0}\right).\label{eq:l-model}
\end{equation}
Here, $\gamma$ is a material-dependent constant. 
Thus a system that is held at a constant strain will evolve to reduce the
stresses in all bonds until a new mechanical equilibrium is reached with a volume determined by the imposed strain. We assume that
the elastic response of the system is much faster than the evolution
of $\ell_{i,0}$. % so that the system is always at force balance. 
This is the dynamics for creep in the Maxwell model of a viscoelastic solid~\cite{maxwell1867iv}.
(Each bond consists of a spring, which describes the rapid elastic behavior, in series with a dash-pot, which at
long times accounts for the change in rest lengths of the spring.) Similar models have been used to describe  junction remodelling in  epithelial cells~\cite{staddon2019mechanosensitive,cavanaugh2019rhoa}.

We will restrict our analysis to the case where every bond has the same stiffness, $k_i=k$. The aging rate is then given by $\Gamma=\gamma k$. More generally, one can define the aging rate in terms of the average stiffness: $\Gamma=\gamma \langle k \rangle$.

Since the dynamics reduce the stress, and therefore the system's energy, the aging process is an energy minimization
algorithm: the rate of change of $\ell_{i,0}$ is proportional to the gradient of the elastic energy:

\begin{equation}
\partial_{t}\ell_{i,0}=-\gamma\frac{\partial U}{\partial\ell_{i,0}}. 
\end{equation}
We use this insight to manipulate the energy landscape.

\section*{Energy landscape picture of training by periodic driving }

Our central goal is to train an elastic system so that a specific source strain, $\epsilon_{S}$, results in a predetermined target response, $\epsilon_{T}$.
%The input and output strains may be either global or local. 
An example of a ``global'' response is tuning the Poisson's ratio $\nu$, so that a uniform
uniaxial strain results in a desired (magnitude and sign) strain in the transverse direction with $\epsilon_T=-\nu \epsilon_S$.
An example of a heterogeneous response would be a strain $\epsilon_S$ applied between source nodes producing a desired strain $\epsilon_T=\Delta \epsilon_S$ at a specified distant target location.  

Our strategy consists of manipulating the energy landscape so that it creates a low energy ``valley'' in the desired direction of $\epsilon_S-\epsilon_T$ space. Ideally, the stiffness in the desired direction would be much lower
than in all other directions. In that case an applied strain, which is not necessarily aligned with the soft direction but has some projection onto it, will actuate the system along the valley direction. 

\begin{figure}
\includegraphics[scale=0.5]{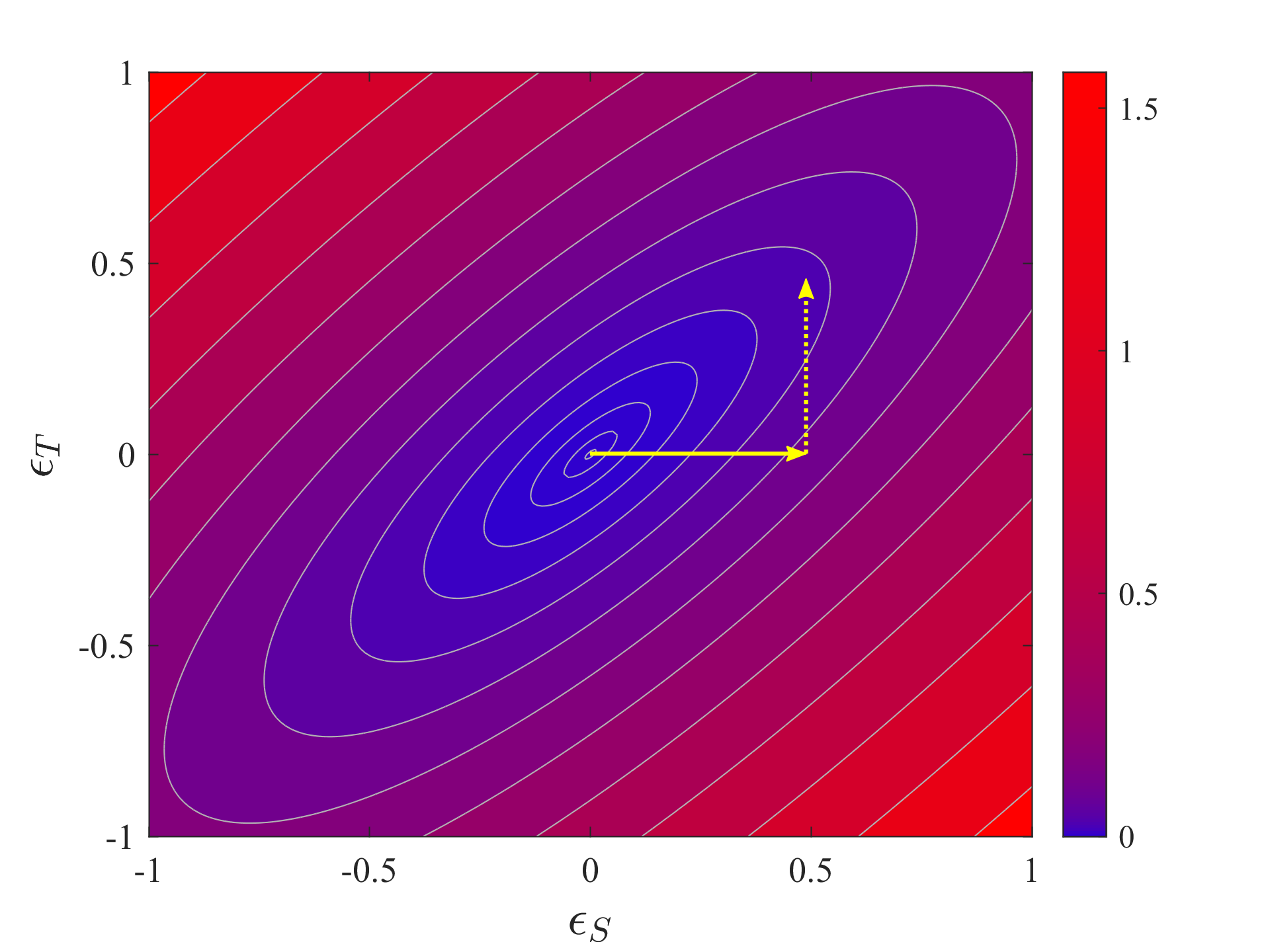}

\caption{An illustration of the energy as function of the source strain and
target strain. The locations of the nodes are defined by force balance. The response is obtained by applying a source
strain (full arrow) $\epsilon_{S}$, and then minimizing the energy
to find $\epsilon_{T}$ (dotted arrows).  \label{fig:Energy}}

\end{figure}

This idea can be illustrated in a simple linear-response model. Consider the energy of a network of $N$ nodes under an applied pair of strains that are fixed 
at $\epsilon_{S}$ and $\epsilon_{T}$, respectively (later on we will remove the constraint
on $\epsilon_{T}$ because the system will be trained to produce the desired $\epsilon_T$ in response to an applied $\epsilon_S$). Under these imposed strains, the remaining unconstrained
nodes evolve to maintain force balance. %Within linear response the locations of the nodes are uniquely determined, but at large strains multiple solutions could exist. 
The elastic energy stored in the network depends on the positions of the $N$ nodes and can be expanded in terms of the two applied strains for small $\epsilon_S$ and $\epsilon_T$:
 %(for simplicity, we set the prefactor of the first two terms to be equal):
\begin{equation}
U=\frac{1}{2}A\epsilon_{S}^{2}+\frac{1}{2}B\epsilon_{T}^{2}+C\epsilon_{S}\epsilon_{T}.\label{eqn:energy}
\end{equation}
To insure that the energy is positive definite we require that $A>0$, $B>0$ and $C^2\leq AB$.
The response can be computed in two steps, as illustrated in Fig.
\ref{fig:Energy}. First, the system is strained by changing $\epsilon_{S}$
while keeping $\epsilon_{T}$ fixed. Then we allow $\epsilon_{T}$
to vary in order to minimize $U$. Requiring that $\left.\frac{\partial U}{\partial\epsilon_{T}}\right|_{\epsilon_{S}}=0$,
leads to 
\begin{equation}
\epsilon_{T}=-\frac{C}{B}\epsilon_{S}.
\end{equation}
When $C\ne0$ the system is anisotropic, and $C/B$ sets the one direction
in which the energy is lower (see Fig. \ref{fig:Energy}). (If $C^2=AB$ then moving along this direction costs zero energy.) % and the output response strain is maximal. 

The aim of our training protocol is to create a valley in the energy landscape of our many-particle system that is similar to the form of Eq.~\ref{eqn:energy}.  This valley will couple $\epsilon_{T}$ to $\epsilon_{S}$ so that the system relaxes to an energy minimum in the $\epsilon_{T}$ direction when $\epsilon_{S}$ is held fixed. By appropriate aging of the system, the aim is to tune $C$ and $B$ so that this minimum will be at the desired value of $\epsilon_{T}$. 

%If the energy landscape is identically zero along the curve $\epsilon_{out}\left(\epsilon_{in}\right)$, then varying $\epsilon_{in}$ while minimizing the energy to find $\epsilon_{out}$, will keep the system along the bottom of this curve.

The energy landscape landscape can be manipulated by straining the
system while the system evolves. As noted, if the system is held at
a given strain this reduces the energy at that strain. To obtain a
range of strains where the energy is low, that is  to create a valley in the energy landscape, we strain the system periodically
along the path $\epsilon_{T}\left(\epsilon_{S}\right)$ . If the system
is strained at a rate which is fast in comparison to the evolution
of $\ell_{i,o}$, the system will minimize the energy at each point along the strained
path.

\section*{Training global response}

We begin by training an auxetic response, defined in terms of the
Poisson's ratio $\nu$. To measure it, we apply a uniaxial strain, $\epsilon_{S}$, along the $x$ direction and measure the resulting transverse strain
along the $y$ direction, $\epsilon_{T}$. The Poisson's ratio is defined
as $\nu=-\frac{\epsilon_{T}}{\epsilon_{S}}$; auxetic materials have $\nu<0$. Within
linear response, for an isotropic $d$-dimensional solid, $\nu<0$ corresponds to a ratio of the shear modulus $G$ to bulk modulus $B$ that satisfies $G/B>d/2$. We will also consider
the non-linear regime where the Poisson's ratio may depend on the magnitude of the imposed strain $\epsilon_{S}$.

To train our networks, we apply periodic strain cycles of isotropic compression and expansion in the continuous range $\left[-\epsilon_{Age},\epsilon_{Age}\right]$.
The goal of this is to reduce the energy along the direction $\epsilon_{T}=\epsilon_{S}$.  We let the system age during the entire cycle so that a smooth continuous valley is created. % If we had only aged the system at $\pm \epsilon_{Age}$ then there could be a significant energy barrier along the valley itself.  

Because the aging rate
increases with the stress in the system, the functional form of the training strain $\epsilon_{tr}$ as a function of time, $\epsilon_{tr}\left(t\right)$,
affects the outcome. Aging is most rapid at strains $\left|\epsilon_{tr}\right|\approx\epsilon_{Age}$
and much slower when $\left|\epsilon_{tr}\right|\approx 0$. To compensate for this effect, we choose
a functional form that emphasizes small strains, $\epsilon_{tr}\left(t\right)=-\epsilon_{Age}\left[g\left(t\right)\right]^{3}$,
where $g\left(t\right)$ is a periodic triangle waveform with unit amplitude
\footnote{Larger exponents age strains more uniformly, however they require
a finer discretization.}. Note that after an integer number of cycles, the energy minimum often differs
from zero strain. It is convenient for $\epsilon=0$ to be the energy
minimum, and therefore at the end of the training phase we also age
the system at zero strain before measuring the elastic properties. 

\begin{figure}
\includegraphics[scale=0.6]{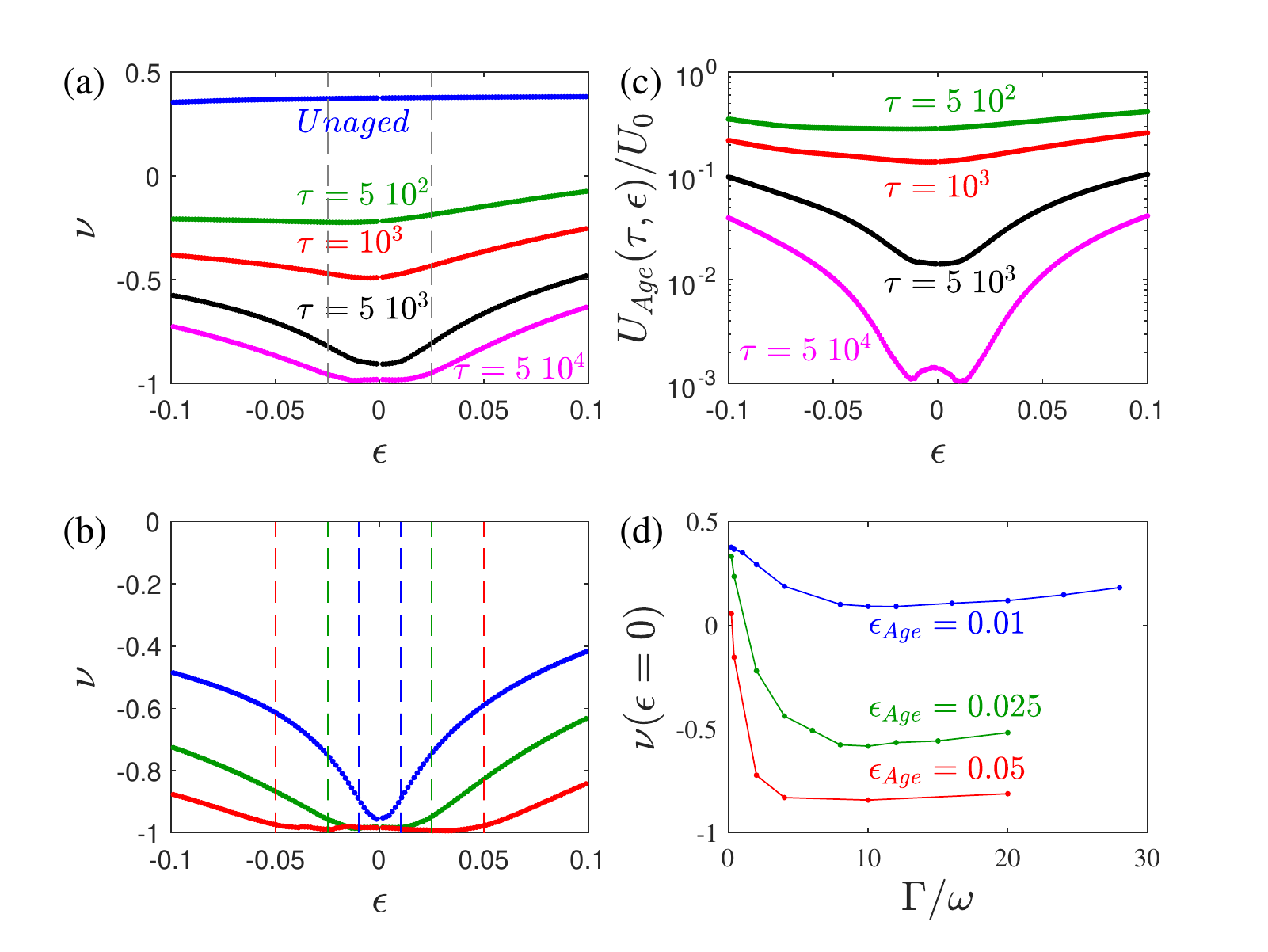}

\caption{ Aging under periodic cycles of isotropic compression and expansion. (a) The Poisson's ratio
$\nu$ versus measurement strain $\epsilon$ for different numbers of cycles, $\tau$. (b) The Poisson's ratio for different maximum aging strains,
$\epsilon_{Age}$, after $\tau=5\cdot10^{4}$ cycles. The Poisson's ratio is near $-1$, in the range $-\epsilon_{Age} \lesssim \epsilon \lesssim \epsilon_{Age}$. Dashed lines denote $\pm\epsilon_{Age}$. (c)
The energy required for isotropic expansion and compression, $U_{Age}(\tau,\epsilon)$ versus isotropic strain $\epsilon$ after $\tau$ cycles, normalized by the
energy of the unaged network. Note the energy is most suppressed at
the limits of the training strain, $\epsilon_{Age}=0.025$.  (d) The linear Poisson's ratio at $\tau=500$ versus the ratio of the aging rate, $\Gamma$,  to the cycle frequency, $\omega$.
\label{fig:Poisson_periodic}}

\end{figure}

In Fig. \ref{fig:Poisson_periodic}(a), we show the evolution of
the Poisson's ratio as a function of the strain at which it is measured $\epsilon$, as the network is aged
for an increasing number of compression/expansion cycles. Initially
the Poisson's ratio is $\nu\approx0.4$ and is only very weakly dependent on strain. As the number of cycles grows,
the Poisson's ratio decreases, especially near zero strain. The Poisson's
ratio is minimal at $\epsilon = 0$ and increases with $\epsilon$ for both positive and negative
strains. For a large number of cycles, $\nu\left(\epsilon\rightarrow0\right)$
approaches $-1$, which is the lowest possible value allowed for
an isotropic solid. 

By aging at different strain amplitudes $\epsilon_{Age}$ we can control
the range over which the Poisson's ratio is minimal. Fig.~\ref{fig:Poisson_periodic}(b) 
shows that $\nu \approx -1$ over the range $-\epsilon_{Age} \lesssim \epsilon \lesssim \epsilon_{Age}$. Thus, the response of the system contains a memory of the range of strains over which the system was aged. Moreover, the response is approximately linear ($\nu$ is nearly independent of $\epsilon$) in this range.
This allows the maximum training strain $\epsilon_{Age}$ to be read out as the strain at which Poisson's ratio starts to depend strongly on strain. These results show that plastic deformation in the $\ell$-model encodes memory stored in the material of the applied strains, as has also been proposed for selecting folding pathways in origami~\cite{stern2019learned}. 

The evolution of the energy required to expand the system to a strain $\epsilon$ as a function of strain
is shown in Fig.~\ref{fig:Poisson_periodic}(c). To emphasize the
relative change at different strains  we normalize the energy, $U\left(\tau,\epsilon\right)$ of the system at strain $\epsilon$ applied after $\tau$ aging cycles,
by the energy before the system was aged, which is approximately quadratic in $\epsilon$.
In the range  $-\epsilon_{Age} \lesssim \epsilon \lesssim \epsilon_{Age}$ the energy is greatly lowered,
decreasing by several orders of magnitude, while at larger strains
the change is more moderate.

We now discuss the optimal amplitude and frequency for periodic
driving in order to direct aging. First, we note that Fig.~\ref{fig:Poisson_periodic}(b) shows that the desired Poisson's ratio $\nu=-1$ is approached at sufficiently high values of the amplitude of the training strain, $\epsilon_{Age}$. This is seen in Fig.~\ref{fig:Poisson_periodic}(d),
which shows the Poisson's ratio at a constant aging time. Increasing
$\epsilon_{Age}$ results in a lower value of $\nu$.  Aging at high strains causes bigger
changes in the structure, which, in turn, have a bigger effect on the elastic
properties. Large strains are
even more important in materials where plasticity occurs only above
a threshold strain.

The next consideration is the frequency of the training strain.
There are two competing scales:
(i) the frequency of the driving, $\omega$, and (ii) the aging rate $\Gamma=\gamma k$ defined below Eq.~\ref{eq:l-model}. Recall that aging is faster when the strains are larger.  The effective aging rate $\Gamma_{eff}$ therefore depends not only on $\Gamma$ but on the aging strain: $\Gamma_{eff}=\Gamma \epsilon_{Age}$. 

If aging occurs at a rate which is high with respect to the frequency of the driving cycle, then the energy will be minimized at each strain during the cycle.
On the other hand if the aging rate is slow then the system will not evolve
much. We therefore, expect that aging is optimal at an intermediate
value, when $\Gamma_{eff} \sim \omega$. In Fig. \ref{fig:Poisson_periodic}(d) we show the Poisson
ratio within linear response versus $\Gamma/\omega \equiv \Gamma_{eff}/\left(\omega \epsilon_{Age}\right)$ at a fixed aging time. In these calculations, $\omega$ is held constant. The effectiveness of aging can be measured by the amount the Poisson's ratio has decreased. The minimum of $\nu$, therefore, is an indication of the optimal aging rate.
 The position of the minimum in $\nu$ shifts to lower values of $\Gamma_{eff}/\left(\omega \epsilon_{Age}\right)$ with increasing $\epsilon_{Age}$, a trend that is consistent with expectation.

%The effect of directed aging on $\nu$ is maximal where $\nu$ itself is at a minimum. The position of the minimum in $\nu$ shifts to lower values of $\Gamma_{eff}/\left(\omega \epsilon_{Age}\right)$ with increasing $\epsilon_{Age}$, a trend that is consistent with expectation.

Aging under periodic isotropic strain is far more effective in producing an auxetic response than aging at a comparable fixed isotropic strain. Fig.~\ref{fig:Poisson_periodic}(b) shows that for a large enough number of cycles, we reach $\nu \approx -1$ over the range of measuring strains of $-\epsilon_{Age} \lesssim \epsilon \lesssim \epsilon_{Age}$ even for small $\epsilon_{Age}$. Aging at a fixed strain of $\epsilon_{Age}$, however, only decreases the Poisson's ratio  under compression~\cite{hexner2019nonlinear}, and by a far smaller amount. At a fixed aging isotropic expansion, $\nu$ actually becomes more positive.  For fixed aging strains, it requires large compressions to become auxetic,  and even then $\nu$ is significantly less negative than for periodic driving. Although the change in response is small over one cycle of driving at small $\epsilon_{Age}$, the changes accumulate over many cycles, eventually leading to dramatic effects.

\begin{figure}
\includegraphics[scale=0.55]{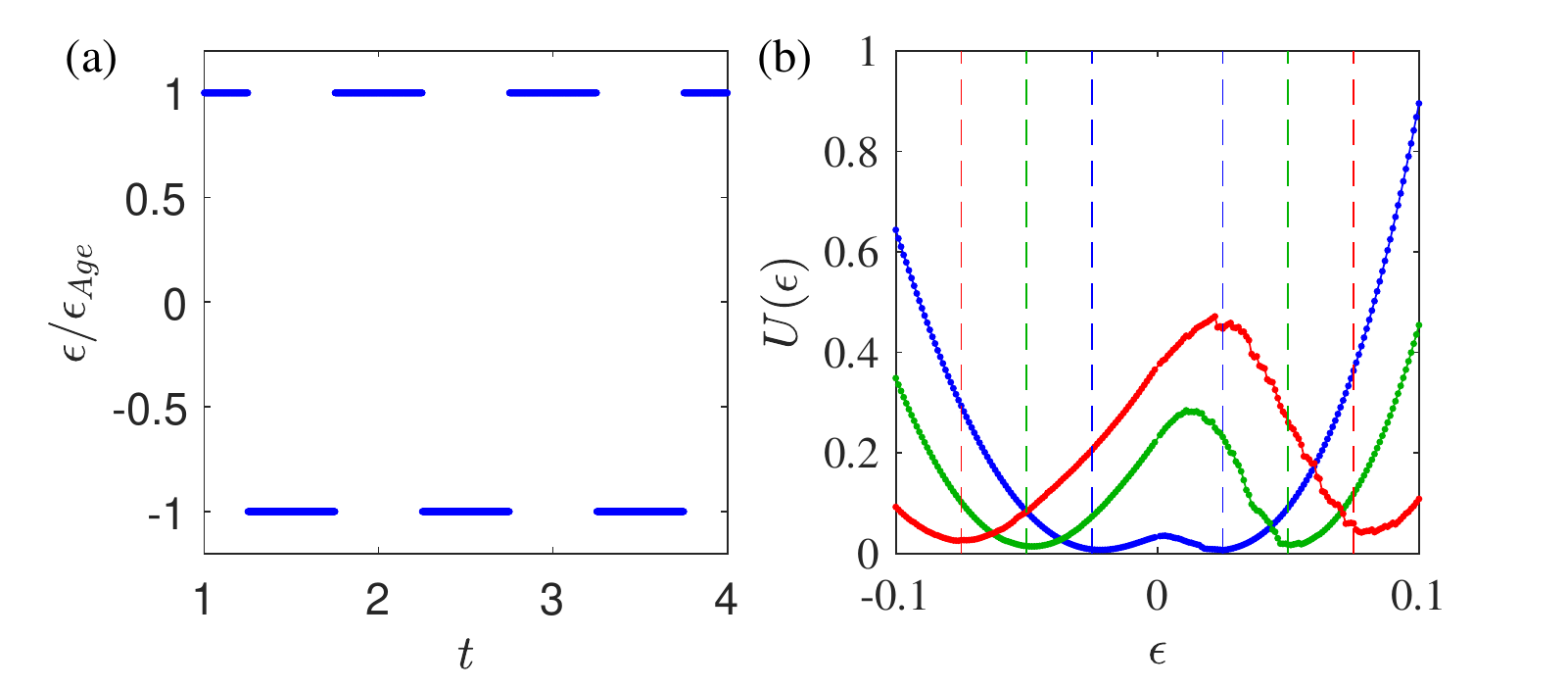}

\caption{Training a bistable energy landscape. (a) The strain as a function of time applied as the system ages. (b) The energy for expansion as a
function of measuring strain after $T=10^4$  cycles of training. The dashed lines denote $\epsilon_{Age}$ and $-\epsilon_{Age}$. 
\label{fig:bistable}}

\end{figure}

We also consider another form for the strain as a function
of time: cyclically switching between $\epsilon=-\epsilon_{Age}$ and
$\epsilon_{Age}$ with a square-wave form, shown in Fig.~\ref{fig:bistable}(a). This lowers
the energy predominantly at the strains $\epsilon_{Age}$ and $-\epsilon_{Age}$. As a result this allows the
system to develop two distinct energy minima, thus allowing bi-stability.
This is demonstrated in Fig.~\ref{fig:bistable}(b). The energy barrier
grows with the amplitude of the training strain. 

Constructing materials with multiple minima allows the creation of materials with discrete states with potentially different elastic properties. Elastic networks with bistable springs have been shown to self-organize in  various ways in response the periodic forcing~\cite{Kedia2019} and have been useful to encode multiple memories~\cite{stern2019learned}.

\section*{Training allosteric response}

\begin{figure}
\begin{centering}
\includegraphics[scale=0.57]{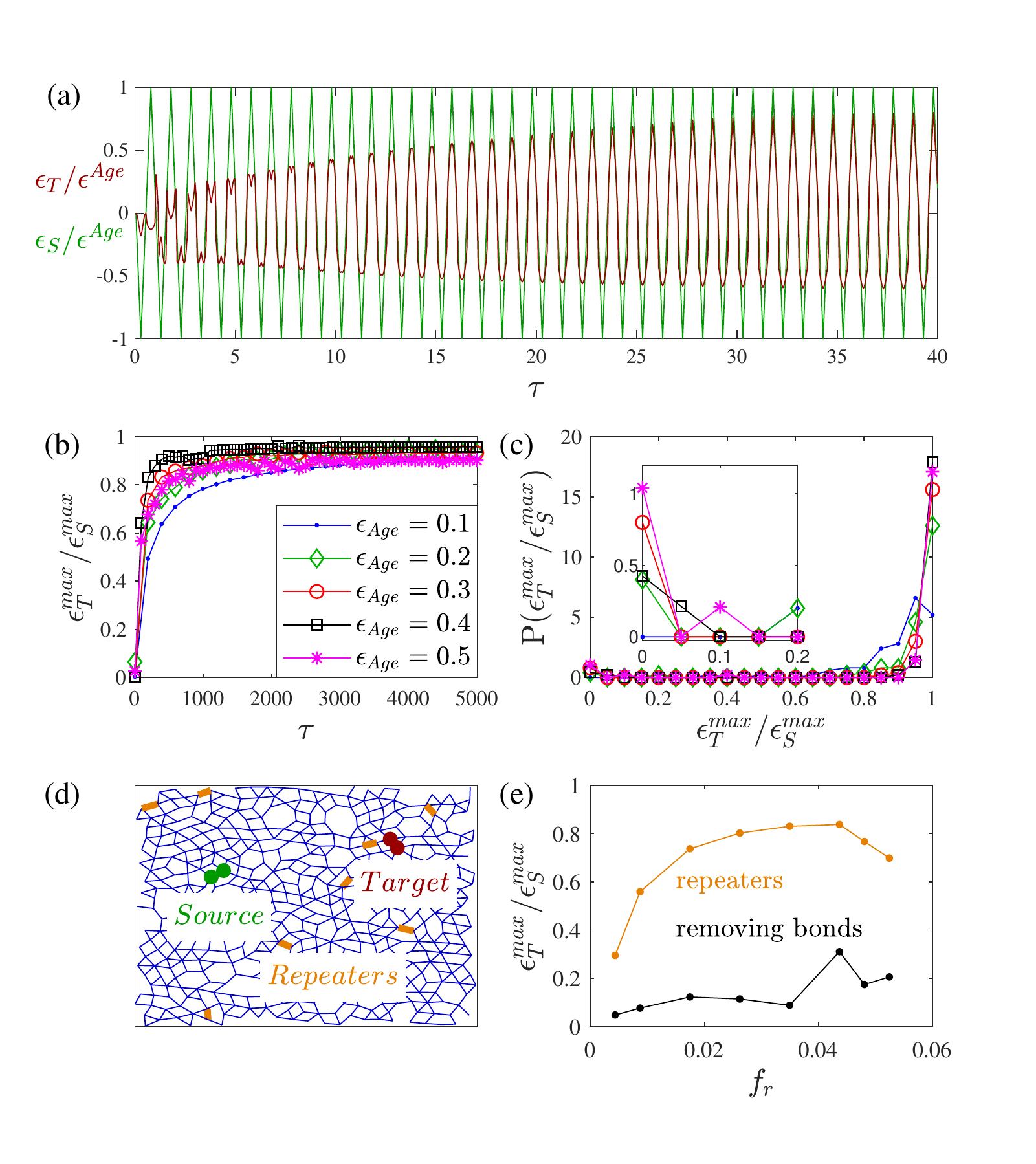}
\par\end{centering}
\caption{(a) The evolution, at short times, of the allosteric response as a function of the number of training cycles $\tau$. After each training cycle we measure the strain on the target nodes (red), $\epsilon_T$ in response to a strain on the source nodes (green), $\epsilon_S$. The maximum value of $\epsilon_S$, $\epsilon^{max}_S$, during the measurement cycle is taken to be the maximum training strain $\epsilon_{Age}$. At large times the maximum value of the target strain, $\epsilon^{max}_T$, converges to its target value: $\epsilon^{max}_T=\epsilon_{Age}$. (b) $\epsilon^{max}_T/\epsilon^{max}_S$ as a function of the number of training cycles $\tau$. Training at near isostatic connectivity is highly successful even far from the linear regime.
(c) The distribution of $\epsilon^{max}_T/\epsilon^{max}_S$ at the largest times measured in (b), using the same color/symbol scheme. Note that the distribution is bimodal: at large $\epsilon_{Age}$ most of the realizations achieve their targets but a small fraction fail nearly completely, with $\epsilon^{max}_T/\epsilon^{max}_S\approx 0$. Inset: a zoomed view of the distribution at small $\epsilon^{max}_T/\epsilon^{max}_S$ shows that more networks fail as $\epsilon_{Age}$ increases. 
(d) An illustration of a network that we train. At large connectivity, we also strain randomly chosen ``repeater'' bonds during training. (e) Training at high connectivity using ``repeaters." For $\Delta Z\approx 0.52$ the optimal number of repeaters is near 4\%.
\label{fig:allostery}}
\end{figure}

We next consider spatially heterogeneous responses, focusing on the
biologically-inspired response of allostery~\cite{rocks2017designing,yan2017architecture} where
the strain on a pair of source nodes results in a strain on a pair
of target nodes. %This can be considered as a ``Greens'' function to an arbitrary deformation. 
For convenience, each pair of source and
target nodes corresponds to the nodes of a randomly chosen bond (that
is subsequently removed). The source and target are chosen to be at least half the system length apart (an example is shown in Fig.~\ref{fig:allostery}(d)).
%The maximal {\bf AJL: don't you mean maximal? NO it should be minimal. The bonds are chosen so that they are far apart} distance between the sourceand target bonds in our system with periodic boundary conditions is half the length of the system (an example is shown in Fig.~\ref{fig:allostery}(d)). 

In the initial unstressed networks the distance between the source nodes
is denoted by $\ell_{S,0}$, and under an imposed strain by $\ell_S$;
the source strain is defined to be $\epsilon_S=\ell_S/\ell_{S,0}-1$. In general, the aim is to produce a target strain $\epsilon_T=\ell_T/\ell_{T,0}-1$ between the target nodes that is given by $\epsilon_T=\Delta \epsilon_S$. For simplicity, we will consider the cases $\Delta=\pm 1$, where squeezing the source
nodes results in a strain between target nodes of the same magnitude, but with a sign that can be chosen to be either positive or negative. 

We begin by training nearly isostatic networks. we strain both the source and target nodes periodically, at a training strain amplitude of $\epsilon_{Age}$, as the system evolves. To produce a target response with $\Delta=1$ the source and training strains are in phase with each other; to produce a target response with $\Delta=-1$, the source and training strains are $\pi$ out of phase. After each cycle of training, we turn off aging to apply one cycle of strain to the source only, and measure the corresponding strain at the target. We then turn the aging back on, train for another cycle and again turn the aging off for another measurement cycle. The results of each measurement cycle are shown in Fig.~\ref{fig:allostery}(a). We show the cycles of source strain $\epsilon_S/\epsilon_{Age}$ in green and the measured target strain $\epsilon_T/\epsilon_{Age}$ in red as a function of the number of training cycles $\tau$ that the system has undergone, for a training designed to produce $\Delta=1$. % in response to an input strain that varies periodically (also with an amplitude $\epsilon_{Age}$). 
Initially, the response at the target is very weak but it grows increasingly stronger as the training continues.  At long times, the response approaches the desired amplitude: $\epsilon_{T} = \epsilon_{S}$.

To characterize the effectiveness of training, we measure $\epsilon^{max}_{T}$, the absolute value of $\epsilon_{T}$ at the largest amplitude of the source strain, $\epsilon^{max}_S=\epsilon_{Age}$. For a completely successful training, $\epsilon^{max}_T/\epsilon^{max}_S=\mid{\Delta}\mid$=1. 
In Fig.~~\ref{fig:allostery}(b) we show  $\epsilon^{max}_{T}/\epsilon^{max}_{S}$ versus the number of training cycles. At long times, the average changes slowly and appears to settle
at a value that depends only weakly on $\epsilon^{max}_{S}$.  At aging strains $\epsilon_{Age}\leq0.5$, on average $\epsilon^{max}_{T}$ reaches above $90\%$  of its target amplitude, $\epsilon^{max}_T=\epsilon_{Age}$.  Training is therefore highly successful even far in the nonlinear regime, producing a target strain that is nearly equal to the source strain even at $\epsilon^{max}_S=0.5$ where the distance between the two source nodes is 50\% higher than the distance in the absence of strain.
 %The response at large strains the is easily trained here, is usually difficult to achieve by design.

%the achieved value of $\epsilon_{out}/\epsilon_{in}$ is on average approximately $0.9$,
Note that the results shown in Fig.~\ref{fig:allostery}(b) correspond to an average over an ensemble of $50$ networks. Fig.~\ref{fig:allostery}(c) shows the distribution of the asymptotic values of $\epsilon^{max}_T/\epsilon^{max}_S$ for the different networks in the ensemble after many cycles of training $\tau$. While most of the networks achieve precisely the targeted response, a fraction of the
realizations fail completely to do so. In the inset of Fig.~\ref{fig:allostery}(c) we show that the fraction of failed realizations grows with $\epsilon_{Age}$.  Our results suggest that a class of low-coordination nodes %(known as ``bucklers'') 
may play an important role in the cases of failure. Networks derived from jammed
packings near the isostatic point, $\Delta Z \approx 0$, are known to have some nodes with only three bonds in two dimensions.
Often two of those bonds meet at angles near $180^{\circ}$ and are thus unable
to support large forces~\cite{hexner2018linking} without buckling; such nodes are known as bucklers.  Squeezing those bonds
results in a localized response~\cite{Charbonneau2015}. We find that if the source and target nodes are allowed to include bucklers, approximately 30\% of networks cannot be tuned at large strain. The analysis showed in Fig.~\ref{fig:allostery}(c) excludes bucklers at the source and target. Even so, approximately 10\% of the networks are untunable, perhaps because bucklers near the source and target nodes inhibit strong responses at high deformations. 

In the nearly isostatic networks discussed so far, we achieve a high degree of success in training allostery by applying strains only to the source and target
nodes while the remaining bonds evolve through plastic deformations. The decay of stresses away from a squeezed bond is governed by an important length scale, $\xi$ that diverges as the coordination is decreased towards isostaticity~\cite{ellenbroek2006critical}.  At distances $r << \xi$ the decay is very slow, and is
almost independent of distance while at $r >> \xi$
there is a crossover to a rapid decay, $r^{-d}$, as expected from
continuum elasticity~\cite{ellenbroek2006critical,lerner2014breakdown}.
When $\xi$ is larger than the distance between the source and
target, the strategy of applying a periodic training strain to the source and target bonds is useful. This is why we achieve a high degree of success in tuning nearly isostatic networks.

However, in more highly coordinated networks, where $\xi$ is small in comparison to the distance between the source and target node pairs, the region in between those two pairs is nearly unaffected by the aging since the strains are negligible. In that case, only bonds near the source and target age significantly, and the source and target never become coupled by the applied strains. 

To overcome this limitation, we introduce applied strains on additional pairs of nodes, which we call ``repeaters''. The goal of the repeaters is to couple the source and target by rebroadcasting
the elastic signal.  Each repeater is a randomly chosen pair of nearby nodes, as illustrated in Fig.~\ref{fig:allostery}(d). 
During the training cycle the repeater nodes are strained periodically with an amplitude $\epsilon_{Age}$, and with a phase, $0$ or $\pi$, chosen randomly.
%just as for the source and target nodes.  Also, l
As with the source and target, the bond between the nodes of each repeater is removed prior to aging. 
Thus, training
does not distinguish between the source, target and repeater nodes. However, during readout, only the source is strained, by $\epsilon_{S}$, while the resulting strain at the target, $\epsilon_{T}$,
is measured. 

Figure~\ref{fig:allostery}(e) shows $\epsilon^{max}_{T}/\epsilon^{max}_{S}$
after the system has reached its asymptotic value for a large number of training cycles, as a function of the fraction
of repeater node pairs. For $\Delta Z\approx0.52$
the response is maximal when $4\%$ of the node pairs are repeaters. This
corresponds to the length scale between repeaters of $\sim5 \langle ell_{0} \rangle$, consistent with the length scale measured in Ref.~\cite{lerner2014breakdown}.
%For a larger number of repeaters. the response is smaller. %possibly since these interfere with each other. 
To ensure that the increase
of $\epsilon^{max}_{T}/\epsilon^{max}_S$ is not the result of bond removal between pairs of repeater nodes,
we also show in Fig.~\ref{fig:allostery}(e) the effect of training when an equivalent number of bonds
are removed at random. This has a much smaller effect.

While our goal is to train the response of a specific target when
the source is strained, the training protocol does not distinguish between
the source, target and the repeaters. This implies that straining
the source nodes also results in an allosteric response in each of the repeaters. Thus, this system achieves multifunctional behavior, where a single
source controls the response at many sites, similar to that studied
in Ref.~\cite{rocks2019limits}. Our results show that it is easier to train multifunctional response in a system in which the targets are spaced approximately a distance $\xi$ apart than it is to train a system with a single target that is at a distance from the source that is large compared to $\xi$. %Interestingly, here training many responses, turns out to be easier than training a single response if the target is far away from the source.

\section*{Discussion}

In this paper we demonstrated that a model that undergoes creep can
be trained to develop unusual mechanical responses well into the non-linear regime.
These can be either global, such as auxetic responses, or spatially heterogeneous, as in a long-ranged allosteric response. Allostery can be considered as a
``Green's function'' that characterizes how a local input strain is transmitted to a far distant local site. Thus, if allostery can be trained into the system then presumably almost
any response can be achieved. %We discuss difficulties in allostery, and propose design strategies. 

The central strategy that we have introduced is based on creating a low-energy valley that couples the source and target strains. This is a curve defined by the target strain as a function of
the source strain $\epsilon_{T}\left(\epsilon_{S}\right)$. In this paper we focused on the simplest case, $\epsilon_{T}\propto \epsilon_{S}$; more generally this curve can be non-linear, and  non-monotonic leading to more complex responses. Straining
periodically along this curve lowers the energy along the entire path. This is a collective effect: driving
a small number of degrees of freedom results in changes to the entire
system. 

This directed aging approach exploits the natural optimization occurring in an aging system. It does not require an initial carefully designed structure nor a careful manipulation
at the microscopic scale during the aging process. 
As a result this method of creating novel function in materials is easily scaled up to systems of arbitrarily large size. %Our results may be validcfor real materials with an Avogadro's number of degrees of freedom.

%Training disordered solids is a problem that is akin to that of learning. 
Our dynamics can be considered a learning
rule by which a system learns a specific motion. The results presented here fit into a broad set of problems in which systems learn by example, such as neural networks. In this class of problems, a large number of variables are
optimized to satisfy a complex constraint. In our system optimization
occurs naturally as the system lowers its energy. This suggests that this is a platform for mechanical machine learning.

We thank Nidhi Pashine, Chukwunonso Arinze, Paul Chaikin and Arvind Murugan for useful discussions. Work
was supported by the NSF MRSEC Program DMR-1420709 (NP), NSF DMR-1404841(SRN) and 
DOE DE-FG02-03ER46088 (DH) and the Simons Foundation for the collaboration
``Cracking the Glass Problem'' awards $\#$348125 to SRN and $\#454945$ to AJL, and Investigator award $\#$327939 to AJL. We acknowledge support from the University of Chicago Research Computing Center.

\bibliography{biblo}
\bibliographystyle{unsrt}

\end{document}